# Revisiting quantum relativistic effects from phase transition by catastrophe theory

Jiu Hui Wu[1], Kejiang Zhou[2] and Shao Kun Yang[1]

[1]*School of Mechanical Engineering, Xi'an Jiaotong University,*
*& State Key Laboratory for Strength and Vibration of Mechanical Structures, Xi'an 710049, China*
[2]*College of Information Science and Engineering, Zhejiang University, Hangzhou 310027, China*

Abstract: In this paper we start from the Schrödinger equation to revisit some classical quantum mechanics from the perspective of phase transition process. Here the relativistic effect of particles moving at high speed can be regarded as the phase transition process when the velocity variable increases. Considering that the catastrophe models could describe qualitatively any phase transition process, we adopt the simplest folding catastrophe type as the potential function in the Schrödinger equation to obtain a revised Schrödinger relativistic equation through the dimensionless analysis first, and then further to derive out the steady-state Klein-Gordon equation and Dirac relativistic equation gradually. These results reveal that the quantum relativistic effect could be considered as the phase transition process, which could be described by adopting the catastrophe models as the potential function in the classical Schrödinger equation.

## Introduction

Due to the fundamentality and complexity of quantum mechanics, its explanation and exploration have always been one of the hotspots in theoretical physics. Jammer has provided a comprehensive classic treatise on fundamental problems and various explanations in quantum mechanics [1].

Dirac proposed that "the present form of quantum mechanics should not be regarded as the final form", it's just "the best theory one can give so far", perhaps in the future "there will be an improved quantum mechanics that brings it back to determinism", but this has to give up some basic ideas that are now considered okay [2].

Classical physics from Newton to General Relativity is essentially the theory of various kinds of smooth behavior. However, there are many kinds of jump phenomena in which sudden changes are caused also by smooth alterations. The sudden changes involved were christened by Thom catastrophes, and the techniques involved to cover a broad range of such phenomena in a coherent manner has become known as catastrophe theory [3]. The catastrophe theory is a highly generalized mathematical theory that summarizes the rules of non-equilibrium phase transition by several catastrophe models, which can explain the phenomenon of gradual quantitative change to sudden qualitative change [4]. In our previous paper, the general non-equilibrium phase transition process of fluid has been investigated quantitatively by using the catastrophe theory [5]. More recently, further we have derived out a novel kind of partial differential equations to link the quantum dynamics and the classical wave motions by adopting the catastrophe models as the potential function in the Schrödinger equation and through the dimensionless analysis [6]. In this paper we will continue to revisit some classical quantum mechanics from the perspective of phase transition processes by using the simplest catastrophe model and starting from the classical Schrödinger equation.

## 1. A revised Schrödinger relativistic equation by catastrophe theory

First we start from the time-dependent Schrödinger equation that can be expressed in this way:

$$i\hbar \frac{\partial \varphi(r,t)}{\partial t} = \left(-\frac{\hbar^2}{2m}\nabla^2 + V(r)\right)\varphi(r,t) \tag{1}$$

where $\nabla$ is the Hamilton's operator, $\varphi$ the wave function, $\hbar$ the Plank's constant, $m$ the rest mass of the particle, and $V(r)$ is the three-dimensional potential function.

To derive out the Dirac relativistic equation from the Schrödinger equation, the relativistic effect of electrons moving at high speed can be regarded as the phase transition process when the velocity variable increases. Therefore the potential function $V(r)$ in Eq. (1) could adopt one of the 7 types of elementary catastrophe models [3, 6].

For simplicity and without losing generality, by adopting the folding catastrophe type $V(x) = x^3 + nx$ (where $x$ denotes the variable and $n$ is the control parameter), the potential function in Eq. (1) could be expressed as $V(r) = \frac{\hbar^2}{2m}[x^3 + n(r)x]$, and the variable $x$ might be expanded to the form of power exponent product with respect to the related parameters as

$$x \sim m^{\alpha_1}\hbar^{\alpha_2}c^{\alpha_3}\omega^{\alpha_4} \tag{2}$$

where $c$ is the speed of light in vacuum, $\omega$ is the circular frequency, and all the indices $\alpha_1$, $\alpha_2$, $\alpha_3$ and $\alpha_4$ are constants to be determined by the non-dimensional analysis. In addition, the parameter $n(r)$ might be taken as $n \sim r^{\alpha_0}$ ($\alpha_0$ is also a constant).

By introducing the three basic dimensions $T$ [s], $L$ [m], and $M$ [kg], it is noted that the dimension of $m$ is $[M]$, the dimensions of $\hbar$, $\omega$ and $c$ are $[ML^2T^{-1}]$, $[T^{-1}]$, $[LT^{-1}]$, respectively. It is noted that the dimension of $x$ is the same as $[L^{-2/3}]$. Thus the relationship among the power exponents by the dimensionless analysis are listed in Table 1.

According to the dimensionless analysis, the power exponents should satisfy the following relationships

$$\begin{cases} 2\alpha_2 + \alpha_3 = -2/3 \\ -\alpha_2 - \alpha_3 - \alpha_4 = 0 \\ \alpha_1 + \alpha_2 = 0 \end{cases} \tag{3}$$

Table 1. Relationship among the power exponents

|   | $m(\alpha_1)$ | $\hbar\ (\alpha_2)$ | $c\ (\alpha_3)$ | $\omega(\alpha_4)$ | $x$ |
|---|---|---|---|---|---|
| L | 0 | 2 | 1 | 0 | $-2/3$ |
| T | 0 | -1 | -1 | -1 | 0 |
| M | 1 | 1 | 0 | 0 | 0 |

By solving the above equation set, we can get that $\alpha_1 = 2/3 - \alpha_4$, $\alpha_2 = \alpha_4 - 2/3$, $\alpha_3 = 2/3 - 2\alpha_4$. Additionally since the dimension of $n(r)x$ is the same as $[L^{-2}]$, there is $\alpha_0 - 2/3 = -2$, from which it is also obtained that $\alpha_0 = -4/3$. Thus the potential function $V(r)$ can be expressed as

$$V(\mathrm{r}) = \frac{\hbar^2}{m}\left[\beta\left(\frac{mc}{\hbar}\right)^2\left(\frac{\hbar\omega}{mc^2}\right)^{3\alpha_4} + B\left(\frac{mc}{\hbar}\right)^{2/3}\left(\frac{\hbar\omega}{mc^2}\right)^{\alpha_4} r^{-4/3}\right] \quad (4a)$$

where $\beta$ and $B$ are both constants.

Let the index $\alpha_4 = 2/3$, due to $E = \hbar\omega$, then Eq.(4a) becomes

$$V(\mathrm{r}) = \beta\frac{E^2}{mc^2} + B(\hbar^2 E/c)^{2/3} r^{-4/3}/m \quad (4b)$$

Here we should notice that, although Eq. (4b) is obtained by the folding catastrophe type, for other catastrophe types adopted as the potential function in Eq. (1), the first term is the same and the second term is in the form of different power exponents of $r$ in Eq. (4b), which also shows the phase transition process of the different levels and from different visions.

Substituting Eq. (4b) into Eq. (1) with rearrangement, we obtain the following time-dependent quantum relativistic field equation

$$i\hbar\frac{\partial\varphi(r,t)}{\partial t} = \left[-\frac{\hbar^2}{2m}\nabla^2 + \beta\frac{E^2}{mc^2} + f(r)\right]\varphi(r,t) \quad (5)$$

where $f(r) = B(\hbar^2 E/c)^{2/3} r^{-4/3}/m$ is the potential function under the action fields. Further considering the quality-energy relationship $E^2 = c^2 p^2 + m^2 c^4$ ($p$ is the momentum), $E^2/(mc^2) = mc^2\left[1 + c^2 p^2/(m^2 c^4)\right]$ denotes the degree of relativistic change in the phase transition process.

In the following from Eq. (5) we will derive out the steady-state Klein-Gordon equation and Dirac relativistic equation gradually.

## 2. The steady-state Klein-Gordon equation and Dirac equation derived

For the relativistic free particles, Eq. (5b) is in this way:

$$i\hbar \frac{\partial \varphi(r,t)}{\partial t} = \left[ -\frac{\hbar^2}{2m} \nabla^2 + \beta \frac{E^2}{mc^2} \right] \varphi(r,t) \tag{6}$$

According to Eq. (6), $\varphi(r,t)$ satisfies that

$$2mc^2 i\hbar \frac{\partial \varphi(r,t)}{\partial t} = [-\hbar^2 c^2 \nabla^2 + 2\beta E^2] \varphi(r,t) \tag{7}$$

For the plane wave $\varphi(r,t) = \varphi(r)\exp(-iEt/\hbar)$, from Eq. (7) there is

$$E^2 \varphi(r) = [-\hbar^2 c^2 \nabla^2 + (2\beta + 1)E^2 - 2mc^2 E]\varphi(r) \tag{8a}$$

Then by rearranging Eq. (8a), we have

$$E^2 \varphi(r) = \{-\hbar^2 c^2 \nabla^2 + [(2\beta + 1)E^2 - 2mc^2 E - m^2 c^4] + m^2 c^4\}\varphi(r) \tag{8b}$$

When $(2\beta + 1)E^2 - 2mc^2 E - m^2 c^4 = 0$, i.e.,

$$\beta = (m^2 c^4 + 2mc^2 E - E^2)/(2E^2) \tag{9}$$

Eq. (8a) becomes

$$E^2 \varphi(r) = \{-\hbar^2 c^2 \nabla^2 + m^2 c^4\}\varphi(r) \tag{10}$$

which is the steady-state Klein-Gordon equation.

Furthermore, let $\hat{H} = i\hbar c \left( S_1 \frac{\partial}{\partial x} + S_2 \frac{\partial}{\partial y} + S_3 \frac{\partial}{\partial z} \right) + S_0 mc^2$, where $S_0$, $S_1$, $S_2$ and $S_3$ are all the operators, if the following conditions are satisfied

$$S_i S_j + S_j S_i = 2\delta_{ij} \quad (i,j = 1, 2, 3) \tag{11}$$

$$S_i S_0 + S_0 S_i = 0 \quad (i = 1, 2, 3) \tag{12}$$

then with $S_0^2 = S_1^2 = S_2^2 = S_3^2 = I$ there is

$$\hat{H}^2 = -\hbar^2 c^2 \nabla^2 + m^2 c^4 \tag{13}$$

Here $\boldsymbol{S} = S_1 \boldsymbol{i} + S_2 \boldsymbol{j} + S_3 \boldsymbol{k}$ and $S_0$ must be 4x4 square Hermit matrices to ensure $\hat{H}$ is a Hermit operator, which can be expressed as

$$S_0 = \begin{bmatrix} I & 0 \\ 0 & -I \end{bmatrix} \quad \text{and} \quad S_j = \begin{bmatrix} 0 & \sigma_j \\ \sigma_j & 0 \end{bmatrix} \quad (j = 1, 2, 3) \tag{14}$$

where $I$ is the unit matrix, and $\boldsymbol{\sigma}$ is the Pauli matrix.

Therefore from Eqs. (10) and (13) we have

$$E\varphi(r) = (i\hbar c \boldsymbol{S} \cdot \nabla + S_0 mc^2)\varphi(r) \tag{15}$$

which is the steady-state Dirac relativistic equation. Further from Eq. (11) it is noted that the angular momentum operator $\boldsymbol{J} = \frac{\hbar}{2}\boldsymbol{S}$ with the reciprocal relationship $\boldsymbol{J} \times \boldsymbol{J} = i\hbar \boldsymbol{J}$, which can describe the electron spin property.

Substitute Eq. (9) into Eq. (5), we can obtain the revised Schrödinger relativistic equation as

$$i\hbar \frac{\partial \varphi(r,t)}{\partial t} = \left[-\frac{\hbar^2}{2m}\nabla^2 + \frac{1}{2}\left(1 + \frac{2E}{mc^2} - \frac{E^2}{m^2c^4}\right)mc^2 + f(r)\right]\varphi(r,t) \qquad (16)$$

which could describe the quantum relativistic effect of particles.

In the near future, we will use the above revised Schrödinger relativistic equation to analyze the magnetic interaction inside the atomic system by the electron spin as well as the physical meaning of the fine structure constant.

## 3. Conclusions

The folding catastrophe model is adopted as the potential function of the Schrödinger equation, from which a revised Schrödinger relativistic equation through the dimensionless analysis is derived out. Furthermore, the steady-state Klein-Gordon equation and Dirac relativistic equation are gradually obtained. These results reveal that the quantum relativistic effect could be considered as the phase transition process, and the potential function in the classical Schrödinger equation may adopt the catastrophe models to describe the quantum relativistic effects.

## References:

[1] Jammer M. The Philosophy of Quantum Mechanics［M］. New York: John Wiley，1974.
[2] Dirac PAM．Development of the physicist's conception of nature［C］. The Physicist's Conception of Nature．(Mehra J． Ed．) D. Reidel Publishing Company，1973．1-14．
[3] Tim Poston and Ian Stewart: Catastrophe Theory and its Applications [M]. London: Pitman, 1978.
[4] Zeeman E. C.: Catastrophe Theory. In: Güttinger W., Eikemeier H. (eds.) Structural Stability in Physics. Springer Series in Synergetics, vol. 4, pp. 12-22. Springer, Berlin, Heidelberg (1979).
[5] Liang Xiao, Wu Jiu Hui, Zhong H. B.: Quantitative analysis of non-equilibrium phase transition process by the catastrophe theory. *Physics of Fluids* 29, 085108 (2017).
[6] Jiu Hui Wu, Lin Zhang, and Kejiang Zhou: A novel kind of equations linking the quantum dynamics and the classical wave motions based on the catastrophe theory, *Europhysics Letters*, Vol.136, No.4, 40004 (2021).